\documentclass[lettersize,journal]{IEEEtran}
\usepackage[T1]{fontenc}

\usepackage[none]{hyphenat}
\usepackage[utf8]{inputenc}
\usepackage[english]{babel}

\usepackage{verbatim}
\usepackage{graphicx}
\usepackage{float}
\usepackage{hyperref}
\usepackage{lipsum}
\usepackage{algorithmic}
\usepackage{algorithm}
\usepackage{textcomp}
\usepackage{stfloats}
\usepackage{url}
\usepackage{amsmath,amsthm}
\usepackage{amssymb}
\usepackage{mathtools}
\usepackage{epstopdf}
\usepackage{booktabs,dcolumn}
\usepackage{multirow}
\usepackage{caption}
\usepackage{subcaption}
\usepackage{ragged2e}
\usepackage{cite}
\usepackage{censor}

\usepackage{array}
\newcolumntype{C}[1]{>{\centering\arraybackslash}p{#1}}
\newcolumntype{L}[1]{>{\RaggedRight\arraybackslash}p{#1}}
\usepackage{wasysym}

\newcommand*\diff{\mathop{}\!\mathrm{d}}

\usepackage{mathrsfs}
\usepackage{xcolor}
\usepackage{colortbl}

\usepackage{caption,hypcap}

\usepackage{cleveref}
\usepackage{scalerel}
\usepackage{tikz}
\usetikzlibrary{svg.path, decorations.markings}

\usepackage{physics}
\usepackage{ifthen}
\usepackage[outline]{contour} 
\usetikzlibrary{calc} 
\usetikzlibrary{angles,quotes} 
\usetikzlibrary{patterns}

\usepackage{soul}

\definecolor{orcidlogocol}{HTML}{A6CE39}
\tikzset{
  orcidlogo/.pic={
    \fill[orcidlogocol] svg{M256,128c0,70.7-57.3,128-128,128C57.3,256,0,198.7,0,128C0,57.3,57.3,0,128,0C198.7,0,256,57.3,256,128z};
    \fill[white] svg{M86.3,186.2H70.9V79.1h15.4v48.4V186.2z}
                 svg{M108.9,79.1h41.6c39.6,0,57,28.3,57,53.6c0,27.5-21.5,53.6-56.8,53.6h-41.8V79.1z M124.3,172.4h24.5c34.9,0,42.9-26.5,42.9-39.7c0-21.5-13.7-39.7-43.7-39.7h-23.7V172.4z}
                 svg{M88.7,56.8c0,5.5-4.5,10.1-10.1,10.1c-5.6,0-10.1-4.6-10.1-10.1c0-5.6,4.5-10.1,10.1-10.1C84.2,46.7,88.7,51.3,88.7,56.8z};
  }
}

\newcommand\orcidicon[1]{\href{https://orcid.org/#1}{\mbox{\scalerel*{
\begin{tikzpicture}[yscale=-1,transform shape]
\pic{orcidlogo};
\end{tikzpicture}
}{|}}}}

\usepackage{suffix}

\DeclarePairedDelimiterX\MeijerM[3]{\lparen}{\rparen}%
{#3\delimsize\vert\,\begin{matrix}#1 \\ #2\end{matrix}}

\newcommand\MeijerG[8][]{%
	G^{\,#2,#3}_{#4,#5}\MeijerM[#1]{#6}{#7}{#8}}

\WithSuffix\newcommand\MeijerG*[7]{%
	G^{\,#1,#2}_{#3,#4}\MeijerM*{#5}{#6}{#7}}

\hypersetup{linkcolor=black,citecolor=black}

\begin{document}

\title{On the Simulation and Correlation Properties of TWDP Fading Process}

\author{Almir~Maric\textsuperscript{\orcidicon{0000-0001-5912-2967}}~\IEEEmembership{}
        and~Pamela~Njemcevic\textsuperscript{\orcidicon{0000-0002-3005-3934}}~\IEEEmembership{}
\thanks{This work was founded by Ministry of Education, Science and Youth of Sarajevo Canton, Bosnia and Herzegovina (grant number)}
\thanks{A. Maric and P. Njemcevic are with Department of Telecommunications, Faculty of Electrical Engineering, University of Sarajevo, Sarajevo, B\&H.}
\thanks{e-mail:\{almir.maric,pamela.njemcevic\}@etf.unsa.ba}
}


\IEEEpubid{\begin{tabular}[t]{@{}c@{}}This work has been submitted for possible publication.\\Copyright may be transferred without notice, after which this version may no longer be accessible.\end{tabular}}

\maketitle

\begin{abstract}
\label{sec:abs}
This paper introduces a novel statistical simulator designed to model propagation in two-way diffuse power (TWDP) fading channels. The simulator employs two zero-mean stochastic sinusoids to simulate specular components, while a sum of sinusoids is used to model the diffuse one. Using the developed simulator, the autocorrelation and cross-correlation functions of the quadrature components, as well as the autocorrelation of the complex and squared envelope, are derived for the first time in literature for channels experiencing TWDP fading. The statistical properties of the proposed simulator are thoroughly validated through extensive simulations, which closely align with the theoretical results.
\end{abstract}


\section{INTRODUCTION}
Mobile radio channel simulators are widely used as cost-effective alternatives to field trials, providing reproducible solutions for rapid performance evaluation by accurately replicating the statistical properties of the real-world communication channels~\cite{Xia06}. Consequently, numerous simulators are relying on filtering or sum-of-sinusoids (SoS) approaches to model various channel conditions.

Among these, the earliest SoS simulators were designed for frequency non-selective Rayleigh channels~\cite{Jak94, Pop01, Zhe02, Xia06, Ali09, Wan12}. They are commonly used to model non-line-of-sight (NLOS) propagation conditions, where no strong specular components exist between the transmitter and the receiver. As a result, such signals consist solely of a diffuse component composed of many low-power scattered waves. Consequentially, corresponding simulators are created by representing the received signal as a superposition of a finite number of sinusoids, mostly with random amplitudes, frequencies, and/or phases. Thereby, to capture the statistical properties of signals received under the described conditions, Clark introduced a mathematical model assuming that the receiver receives $N \to \infty$ scattered waves with equal amplitudes, but with random angles of arrival (AoA) and random phases, mutually independent and uniformly distributed over $[-\pi,\pi)$~\cite{Cla68}. Clark's model has been validated through numerous measurements~\cite{Ski14} and has become a reference model for modeling propagation in NLOS conditions. As such, it has been used as the foundation of many Rayleigh SoS simulators, which, based on different assumptions about the behavior of sinusoidal phases and AoAs, aim to reproduce statistical properties of the Clark’s model, simultaneously maximizing computational efficiency by minimizing the number of employed sinusoids~\cite{Ski14}.

The simulators mentioned above are also employed to model low-power scattered waves within Rician simulators, which are designed to simulate channels where the received signal consists of one strong specular component (typically line-of-sight, LoS) alongside a diffuse component.  As a result, many Rician SoS simulators have been proposed to date, each exhibiting more or less desired statistical properties and specific computational efficiencies~\cite{Xia06, Ali09,Wan12}. 

However, to date, no attention has been given to the development of simulators 
that considers more than one specular component, such as the TWDP channel. In these channels, the received signal consists of two dominant specular components along with numerous low-power scattered waves, and as a result, its first- and second-order statistics differ significantly from those in Rayleigh or Rician channels. 

Consequently, this paper presents a SoS-based simulator for modeling propagation in TWDP channels. In this simulator, the specular components are modeled as zero-mean stochastic sinusoids with pre-determined angles of arrival and random initial phases, while the diffuse component is simulated using the NLOS model proposed in~\cite{Xia06}. The simulator is used to derive the autocorrelation and cross-correlation functions of the quadrature components, as well as the autocorrelation of the complex envelope and its square, for signals propagating in TWDP channels. The statistical properties of the proposed simulator are compared with those calculated for the reference model, demonstrating excellent agreement between them.

\section{The reference TWDP channel model}
TWDP model assumes that the normalized envelope of the received signal, consisted of two specular and diffuse components, can be expressed as~\cite{Dur02}:
\begin{equation} 
\label{z(t)1}
z(t)=\frac{1}{\sqrt{\Omega}}\left(V_1e^{j\phi_1(t)}+V_2e^{j\phi_2(t)}+n(t)\right)
\end{equation}
where $V_1$ and $V_2$ are magnitudes of the specular components which remain constant over the local stationary interval, $\phi_1(t)$ and $\phi_2(t)$ are random phases of specular components, $n(t)$ is the zero-mean Gaussian-distributed diffuse component with an average power equals to $2\sigma^2$, and $\Omega=V_1^2+V_2^2+2\sigma^2$ is the average power of the overall TWDP fading process. 

\begin{figure}[h]
	\centering
	\begin{tikzpicture} [scale=1.8,every node/.style={scale=0.9},decoration={markings,    mark=at position 0.55 with {\arrow{>}}}]

        \tikzstyle{ground}=[preaction={fill,top color=black!10,bottom color=black!5,shading angle=30},fill,pattern=north east lines,draw=none,minimum width=0.3,minimum height=0.6]
        \def\W{0.7} 
        \def\D{0.1} 

	\draw [very thick,rotate=-20] (-2.,1.5) -- (-1.3,1.5);
        \draw[ground, rotate=-20] (-2,1.5) rectangle++ (\W,\D);
        \draw [thick, postaction={decorate}, blue] (-2.0,1.6) -- (-1.8,0.3);
        \draw [thick, postaction={decorate}, blue] (-1.8,0.3) -- (0,0);
 	\draw [very thick,rotate=-50] (-1.7,-1.2) -- (-1.0,-1.2);
        \draw[ground, rotate=-50] (-1.7,-1.2) rectangle++ (\W,-\D);
        \draw [thick, postaction={decorate}, red] (-1.08,1.98) -- (-0.0,0.0);
        \draw [thick, postaction={decorate}, red] (-2.0,1.6) -- (-1.08,1.98);
        
        \draw (-2,1.6) node[above left] {$T_x$};
	\filldraw[very thick] (0cm+1pt,0cm-1pt) rectangle (0cm-1pt,0cm+1pt);

        \draw (0,0) node[below ] {$R_x(t)$};

        
        \filldraw[very thick] (0.7cm+1pt,0.3cm-1pt) rectangle (0.7cm-1pt,0.3cm+1pt);
        \draw (0.7,0.3) node[right] {$R_x(t+\Delta t)$};
        \draw [thick, dashed, postaction={decorate}, blue] (-2.0,1.6) -- (-1.95,0.47);
        \draw [thick, dashed, postaction={decorate}, blue] (-1.95,0.47) -- (0.7,0.3);
        \draw [thick, dashed, postaction={decorate}, red] (-2.0,1.6) -- (-0.93,1.92);
        \draw [thick, dashed, postaction={decorate}, red] (-0.93,1.92) -- (0.7,0.3);

        \draw [thin,blue] (0.08,0.34) arc (0:-90:0.05);
        \draw [thin, densely dotted, blue] (0,0) -- (0.04,0.35);
        \draw [thin, densely dotted, red] (0,0) -- (0.52,0.48);
        \draw [thin, red, rotate=-40] (0.13,0.70) arc (0:-90:0.05);
        \draw (0.73,0.53) node[rotate=-46, text=red] {$\Delta l_2$};
        \draw [thin, red, <->] (0.55,0.53) -- (0.74,0.34);
        \draw (0.3,0.49) node[rotate=-5, text=blue] {$\Delta l_1$};
        \draw [thin, blue, <->, rotate=-5] (0.0,0.42) -- (0.67,0.42);

        \shadedraw[inner color=red!80, outer color=red!20, draw=black] (-2.0,1.6) circle (1.5pt); 
			
	\draw [thin,densely dotted] (0cm,0cm) -- (1.6cm,0.7cm);
	\draw [thin, -latex, blue, rotate=23] (0.4,0) arc (0:148:0.4);
	\draw (-0.47,0.0) node[ text=blue] {$\alpha_1$};	
	\draw [thin, -latex, red, rotate=23] (0.2,0) arc (0:99:0.2);
	\draw (0.07,0.) node[right, text=red] {$\alpha_2$};
	\draw (2,2) node[above right] {$\mathbb{R}^2$};

        \draw [ultra thick, rotate=23.5, -latex] ((0,0) -- (0.6,0.0) node[below] {$\vec{v}$};
	\end{tikzpicture}
	\caption{Channel model - illustration of specular reflections}
	\label{3_fig003}
\end{figure}
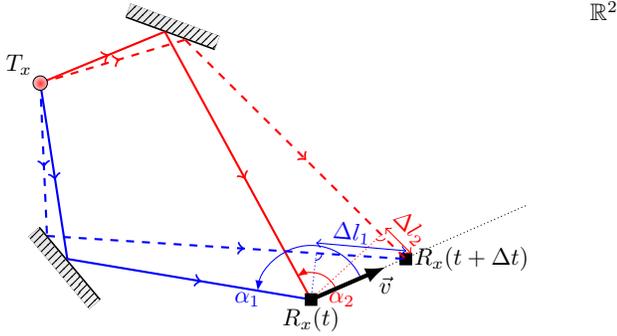

\IEEEpubidadjcol In order to simulate the above defined TWDP fading process, let's assume communication scenario in which the receiver moves with constant velocity vector $\vec{v}$ (as illustrated 

\noindent in Fig.~\ref{3_fig003}) and in which the signal propagates from the transmitter to the receiver via two single-bounce specular reflections, as well as diffuse reflections from local scatterers in the vicinity of the receiver. 

Due to the motion of the receiver, the instantaneous phase of each specular component's phase, $\phi_i(t)$ for $i=\overline{1,2}$, changes in time as:
\begin{equation}
\label{fi_i1}
        \phi_i(t)=2\pi \frac{l_i(t)}{\lambda}     =\frac{2\pi}{\lambda} \sqrt{l_i^2-2 l_i v t \cos{\alpha_i} +\left(v t\right)^2}
\end{equation}
where the angle of arrival $\alpha_i$ can be considered as time invariant only during the  local stationarity interval. Within this interval, the instantaneous phase of each specular component can be approximated as:
\begin{equation}
\label{fi_i2}
    \begin{aligned}
        \phi_i(t)&\approx 2\pi \frac{l_i}{\lambda} - 2\pi \frac{v t}{\lambda} \cos{\alpha_i}\\
        &\approx \phi_i - 2\pi f_D t \cos{(\alpha_i)}\\
        &\approx \phi_i - \Dot{\phi}_i t
    \end{aligned}
\end{equation}
where $\lambda$ is the signal's wavelength, $f_D=v/\lambda$ is maximum Doppler frequency, and $\Dot{\phi}_i=-2\pi f_D \cos{\alpha_i}$ is the velocity of phase change. Thus, the normalized lowpass fading process of the received signal can be described as:
\begin{equation}
\label{z(t)2}
    \begin{aligned}
        z(t)=\frac{1}{\sqrt{\Omega}}\left(V_1 e^{j\left(\phi_1+\Dot{\phi}_1 t\right)}+V_2 e^{j\left(\phi_2+\Dot{\phi}_2 t\right)}+n(t)\right)
    \end{aligned}
\end{equation}
Assuming further that the signal propagates over a wide-sense stationary with uncorrelated scatterers (WSSUS) channel, the reflectors can be treated as independently positioned in 2D space. Consequently, $\phi_1$, $\phi_2$, $\alpha_1$, and $\alpha_2$ are considered as independent and identically distributed (i.i.d.) uniform random variables over the interval $[-\pi,\pi)$. Under these conditions, the specular components can be modeled as zero-mean stochastic sinusoids with predetermined angles of arrival and random initial phases, which accurately capture their physical characteristics~\cite{Xia06}.

Considering the aforementioned and the fact that the specular and scattered components are all zero-mean with mutually independent initial phases, the autocorrelation of the TWDP complex envelope's in-phase component, $x(t)=Re\{z(t)\}$, can be obtained by ensemble averaging, as: 
\begin{equation}
\label{Rxx}
    \begin{aligned}
    \begin{split}
        &R_{xx}(t,t+\tau)=\mathbb{E}\left\{x(t) x(t+\tau)\right\}\\
        &\quad=
         \frac{V_1^2}{2\Omega}\cos(\dot{\phi_1} \tau)
         +\frac{V_2^2}{2\Omega}\cos(\dot{\phi_2} \tau)+\frac{1}{\Omega}
       \mathbb{E}\Bigl\{n_x(t)n_x(t+\tau)\Bigr\}
            \end{split}
    \end{aligned}
\end{equation}
where $\mathbb{E}\{\cdot\}$ is the statistical expectation operator,  $\tau$ is the time lag and
$\mathbb{E}\left\{n_x(t)n_x(t+\tau)\right\}$ is the autocorrelation function of the scattered waves' in-phase component, whose value depends on a chosen NLOS model.
The autocorrelation of the quadrature component $R_{yy}(t,t+\tau)$,  cross-correlations of the in-phase and quadrature  components $R_{xy}(t,t+\tau)$ and $R_{yx}(t,t+\tau)$, and the autocorrelation of the complex envelope $R_{zz}(t,t+\tau)$, can be obtained in a similar manner, and here, the general expressions are not presented for brevity.
Finally, the autocorrelation of the squared envelope, can be expressed as (\ref{Rz2z2}),
\begin{figure*}
\begin{equation}
\label{Rz2z2}
    \begin{split}
        &R_{|z|^2|z|^2}(t,t+\tau)=\mathbb{E}\Bigl\{|z(t)|^2 |z(t+\tau)|^2\Bigr\}\\
        &\qquad=\frac{1}{\Omega^2} \biggl[ \mathbb{E}\Bigl\{n_x^2(t)n_x^2(t+\tau)\Bigr\}+\mathbb{E}\Bigl\{n_y^2(t)n_y^2(t+\tau)\Bigr\}+\mathbb{E}\Bigl\{n_x^2(t)n_y^2(t+\tau)\Bigr\}+\mathbb{E}\Bigl\{n_y^2(t)n_x^2(t+\tau)\Bigr\}\\
        &\qquad+\Bigl(\mathbb{E}\Bigl\{n_x^2(t)\bigr\}+\mathbb{E}\Bigl\{n_y^2(t)\Bigr\}+\mathbb{E}\Bigl\{n_x^2(t+\tau)\Bigr\}
        +\mathbb{E}\Bigl\{n_y^2(t+\tau)\bigr\}\Bigr)\Bigl(V_1^2+V_2^2\Bigr)+\Bigl(V_1^2+V_2^2\Bigr)^2+\\
        &\qquad+2\Bigl(\mathbb{E}\Bigl\{n_x(t)n_x(t+\tau)\Bigr\}+\mathbb{E}\Bigl\{n_y(t)n_y(t+\tau)\Bigr\}\Bigr)\Bigl(V_1^2\cos{\dot{\phi_1}\tau}+V_2^2\cos{\dot{\phi_2}\tau}\Bigr)+        2V_1^2V_2^2\cos({(\dot{\phi_1}-\dot{\phi_2})\tau})+\\
        &\qquad+2\Bigl(\mathbb{E}\Bigl\{n_x(t)n_y(t+\tau)\Bigr\}-\mathbb{E}\Bigl\{n_y(t)n_x(t+\tau)\Bigr\}\Bigr)  \Bigl(V_1^2\sin{\dot{\phi_1}\tau}+V_2^2\sin{\dot{\phi_2}\tau}\Bigr)\biggr]
        \end{split}
\end{equation}
\end{figure*}
where $\mathbb{E}\left\{n_y(t)n_y(t+\tau)\right\}$ is the autocorrelation of the scattered waves' quadrature component, $\mathbb{E}\left\{n_x(t)n_y(t+\tau)\right\}$ and $\mathbb{E}\left\{n_y(t)n_x(t+\tau)\right\}$ are their cross-correlations, while $\mathbb{E}\bigl\{n_x^2(t)n_x^2(t+\tau)\bigr\}+\mathbb{E}\bigl\{n_y^2(t)n_y^2(t+\tau)\bigr\}+\mathbb{E}\bigl\{n_x^2(t)n_y^2(t+\tau)\bigr\}+\mathbb{E}\bigl\{n_y^2(t)n_x^2(t+\tau)\bigr\}$ is the autocorrelation of the scattered waves' squared envelope, all dependent on  underlying model chosen to simulate the behavior of the scattered waves. 

Within the isotropic scattering environment and under the narrow-band ﬂat fading assumption, the diffuse component $n(t)$ can be mathematically modeled using Clark's model, as a superposition of $N$ sinusoids. As such, it can be expressed as~\cite{Xia06}: 
\begin{equation}
    \label{n(t)}
    n(t)=\sqrt{\frac{2\sigma^2}{N}} \sum\limits_{i=1}^{N}  e^{j\left(2\pi f_D t \cos{ \beta_i}+\varphi_i\right)}
\end{equation}
where $N$ is the number of scattered propagation paths, while $\beta_i$ and $\varphi_i$ are the angle of arrival and the initial phase of $i$-th path, respectively, assumed to be mutually independent and uniformly distributed over $[-\pi,\pi)$ for all $i$~\cite{Cla68}. 

So, in case when ideal Clark's model is used to model diffuse component, when $N \to \infty$, the reference expression for autocorrelation of the TWDP complex envelope's in-phase component can be obtained by inserting~\cite[(2b)]{Pat98} into (\ref{Rxx}), as: 
\begin{equation}
\label{Rxx1}
    \begin{aligned}
        R_{xx}(\tau)=\frac{V_1^2}{2\Omega}\cos{\left(\Dot{\phi}_1 \tau\right)}+\frac{V_2^2}{2\Omega} \cos{\left(\Dot{\phi}_2 \tau\right)}+\frac{2\sigma^2}{2\Omega} J_0\left(2 \pi f_D \tau \right)
    \end{aligned}
\end{equation}
where $J_0(\cdot)$ is the zero-order Bessel function of the ﬁrst kind.
Under the same assumptions, the reference expressions for autocorrelation of the quadrature component, the cross-correlations of the in-phase and quadrature  components, as well as the autocorrelations of the complex envelope and its square, can be obtained in a similar manner, using the normalized results given by~\cite[(4a)-(4f)]{Ski14}, as:
\begin{subequations}
\begin{align}
        &R_{yy}(\tau)=R_{xx}(\tau)\label{Ryy1}\\
        &R_{xy}(\tau)=-R_{yx}(\tau)=     \frac{V_1^2}{2\Omega}    \sin{\left(\Dot{\phi}_1 \tau\right)}+\frac{V_2^2}{2\Omega}    \sin{\left(\Dot{\phi}_2 \tau\right)}\label{Rxy1}\\
        &R_{zz}(\tau)
        =
        \frac{V_1^2}{\Omega} e^{-j \dot{\phi_1} \tau}+\frac{V_2^2}{\Omega} e^{-j \dot{\phi_2} \tau}+\frac{2\sigma^2}{\Omega} J_0(2 \pi f_D \tau)\label{Rzz1}\\
        &R_{|z|^2|z|^2}(\tau)=\frac{2\sigma^2}{\Omega}J_0(2\pi f_D \tau)
        \Bigl[\frac{2\sigma^2}{\Omega}J_0(2\pi f_D \tau)\notag\\
        &\qquad+2\frac{V_1^2}{\Omega}   \cos{\bigl(\dot{\phi_1}\tau\bigr)}+2\frac{V_2^2}{\Omega}\cos{\bigl(\dot{\phi_2}\tau\bigr)}\Bigr]\label{Rz2z21}\\
        &\qquad+1+2\frac{V_1^2V_2^2}{\Omega^2}\cos{\left((\dot{\phi_1}-\dot{\phi_2})\tau\right)}\notag
\end{align}
\end{subequations}
Note that the derived expressions represent the statistical correlation properties of non-realizable stochastic reference model for the TWDP channel (since they are derived for $N \to \infty$). Nevertheless, these results are crucial, as they serve as the benchmark for evaluating the performance of any proposed TWDP simulator~\cite{Pat06}, which should aim to reproduce these properties as accurately as possible~\cite{Pat05}.

\section{The TWDP channel simulator}
Based on the described reference model, the TWDP channel simulator is constructed following previously 
derived assumptions about the characteristics of the specular components. However, to simulate diffuse component using finite number of sinusoids, a detailed literature overview is performed, reveling the existence of many different models~\cite{Jak94, Pop01, Zhe02, Xia06, Ali09, Wan12}, etc. Among these, the model proposed by~\cite[(6a)-(6c), (7)]{Xia06} (where $\beta_i$ in (\ref{n(t)}) is defined as $\beta_i=(2 \pi i + \vartheta_i)/N$ and $\vartheta_i$ and $\varphi_i$ are statistically independent and uniformly distributed over $[-\pi,\pi)$) is chosen due to its  favorable simulation time, simplicity and pretty accurate correlation statistics~\cite{Pat05}. So, based on the assumption that the initial phases 
of the specular components are random and  independent from the initial phases of the scattered waves, chosen model is used to obtain autocorrelation and cross-correlation functions of the simulated TWDP signal, as:
\begin{subequations}
\begin{align}
        &R_{xx}(\tau)=R_{yy}(\tau)        =\frac{V_1^2}{2\Omega}\cos{\left(\Dot{\phi}_1 \tau\right)}\label{Rxx2}\\
        &\qquad+\frac{V_2^2}{2\Omega} \cos{\left(\Dot{\phi}_2 \tau\right)}+\frac{2\sigma^2}{2\Omega} J_0\left(2 \pi f_D \tau\right)\notag\\
        &R_{xy}(\tau)=-R_{yx}(\tau)=\frac{V_1^2}{2\Omega}\sin{\left(\Dot{\phi}_1 \tau\right)}+\frac{V_2^2}{2\Omega}\sin{\left(\Dot{\phi}_2 \tau\right)}\label{Rxy2}\\
        &R_{zz}(\tau)=\frac{V_1^2}{\Omega} e^{-j \dot{\phi_1} \tau}+\frac{V_2^2}{\Omega} e^{-j \dot{\phi_2} \tau}+\frac{2\sigma^2}{\Omega} J_0(2 \pi f_D \tau)\label{Rzz2}\\
        &R_{|z|^2|z|^2}(\tau)=\frac{2\sigma^2}{\Omega}J_0(2\pi f_D \tau)
        \Bigl[\frac{2\sigma^2}{\Omega}J_0(2\pi f_D \tau)\notag\\
        &\qquad +2\frac{V_1^2}{\Omega}   \cos{\bigl(\dot{\phi_1}\tau\bigr)}+2\frac{V_2^2}{\Omega}\cos{\bigl(\dot{\phi_2}\tau\bigr)}\Bigr]\label{Rz2z22}\\
        &\qquad+1+2\frac{V_1^2V_2^2}{\Omega^2}\cos{\left((\dot{\phi_1}-\dot{\phi_2})\tau\right)}\notag\\
        &\qquad-\frac{4\sigma^4}{\Omega^2} f_c(2\pi f_D \tau,N)-\frac{4\sigma^4}{\Omega^2}  f_s(2\pi f_D \tau,N)\notag
\end{align}
\end{subequations}
where $f_c(x,N)$ and $f_s(x,N)$ are defined as~\cite{Xia06}:
\begin{equation}
    \begin{split}
        f_c(x,N)=\sum_{n=1}^{N} \left(\frac{1}{2\pi} \int_{\frac{2\pi n-\pi}{N}}^{\frac{2\pi n+\pi}{N}} \cos\left(x \cos\gamma \right) \diff{\gamma}\right)^2
    \end{split}
\end{equation}
\begin{equation}
    \begin{split}
        f_s(x,N)=\sum_{n=1}^{N} \left(\frac{1}{2\pi} \int_{\frac{2\pi n-\pi}{N}}^{\frac{2\pi n+\pi}{N}} \sin\left(x \cos\gamma \right) \diff{\gamma}\right)^2
    \end{split}
\end{equation}

Obviously, the autocorrelation and cross-correlation functions given by (\ref{Rxx2})–(\ref{Rzz2}) do not depend on the number of sinusoids $N$, and they exactly match the desired second-order statistics of the reference TWDP model, given by (\ref{Rxx1}), (\ref{Ryy1})-(\ref{Rzz1}). However, the autocorrelation function of the squared envelope, given by (\ref{Rz2z22}), differs from that calculated for the reference model. Despite this, it asymptotically approaches the desired autocorrelation (\ref{Rz2z21}) as the number of sinusoids approaches infinity, and good approximation can be obtained even for relatively small number of sinusoids (e.g. $N\geq 8$) for most combinations of $V_1$, $V_2$, and $2\sigma^2$. 

The proposed simulator is obviously wide-sense-stationary (WSS), since its mean value is constant and its autocorrelation function depends only on the time difference $\tau$, i.e. $R_{zz}(t,t+\tau)=R_{zz}(\tau)$~\cite{Pat06}. 

By analyzing correlation functions (\ref{Rxx2})-(\ref{Rz2z22}), it can also be shown that the obtained results reduce to those given by~\cite[(12a)-(12d)]{Xia06} and calculated using the Rician fading simulator, in case when $V_2=0$ (for $K=V_1^2/\sigma^2$ and $\Omega=V_1^2+2\sigma^2$).

\section{Numerical results}
Verification of the proposed TWDP fading simulator is performed
by comparing simulation results for finite
$N$ with those of the theoretical limits when $N$ approaches
infinity, for many different combinations of specular components' magnitudes $V_1$, $V_2$ and diffuse component's strength $2\sigma^2$. Thereby, in this section, instead of using specific values of $V_1$, $V_2$ and $2\sigma^2$, common TWDP parameters $K$ and $\Gamma$, defined as: $K=(V_1^2+V_2^2)/(2\sigma^2)$ and $\Gamma=V_2/V_1$, for $V_2\leq V_1$, are used to describe specific propagation conditions in TWDP channels, in order to clearly perceive the impact of different fading severities on TWDP correlation characteristics. 

Accordingly, the simulations are conducted for many different combinations of TWDP fading parameters $K$ and $\Gamma$ and different combinations of AoAs ($\alpha_1$ and $\alpha_2$). Throughout the simulations performed, the number of sinusoids is chosen to be $N = 8$ and all the ensemble averages for the simulation results are based on 500 random trials (as suggested in~\cite{Xia06}). Also, the normalized sampling period and maximum Doppler frequency are chosen to be $f_D T_s=0.01$ (where $T_S$ is the sampling period) and $f_D=1000$ Hz, respectively. 


\subsection{Evaluation of Correlation Statistics}
\begin{figure}[t]
    \vspace{-0.4cm}
	\centering        \includegraphics[width=.425\textwidth]{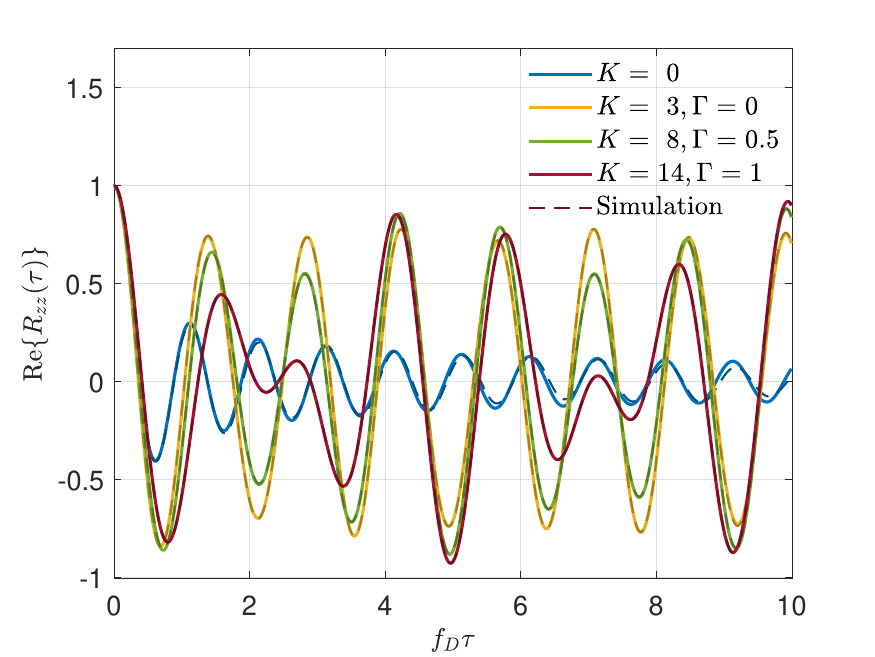}
	\caption{The real part of the autocorrelation of TWDP complex envelope ($\alpha_1=\pi/4$, $\alpha_2=2\pi/3$)}
	\label{Fig_Rxx}
\end{figure}

\begin{figure}[t]
    \vspace{-0.4cm}
	\centering        \includegraphics[width=.425\textwidth]{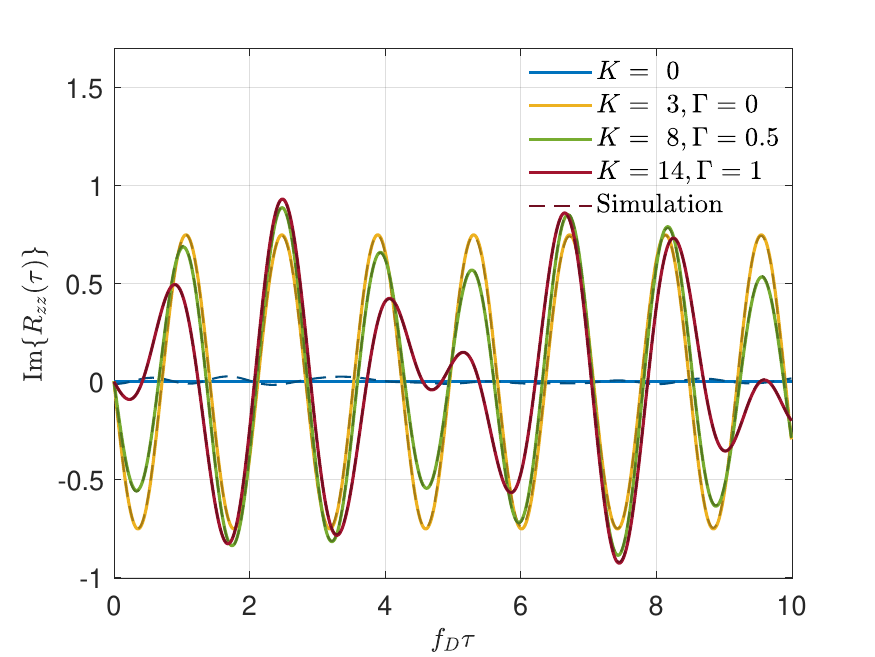}
	\caption{The imaginary part of the autocorrelation of TWDP complex envelope ($\alpha_1=\pi/4$, $\alpha_2=2\pi/3$)}
	\label{Fig_Rxy}
\end{figure}

\begin{figure}[t]
    \vspace{-0.4cm}
	\centering        \includegraphics[width=.425\textwidth]{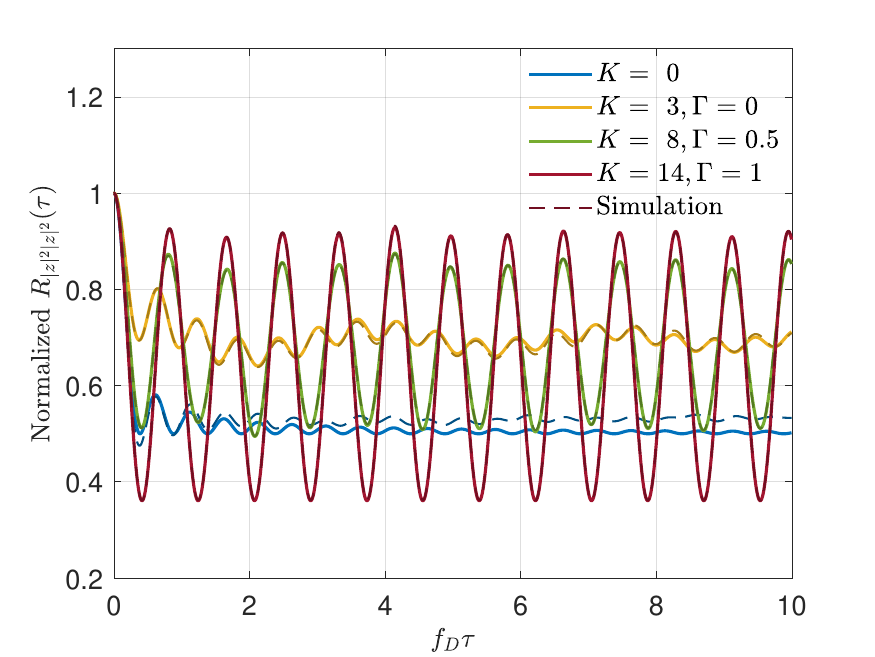}
	\caption{The autocorrelation of TWDP squared envelope ($\alpha_1=\pi/4$, $\alpha_2=2\pi/3$)}
	\label{Fig_Rz2z2}
\end{figure}

Although extensive simulations are performed to evaluate TWDP correlation statistics, due to the space limitations, only simulation results obtained for the real and imaginary part of the autocorrelation of the complex envelope, together with the autocorrelation function of the envelope square,
 are shown in Figs. (\ref{Fig_Rxx}) - (\ref{Fig_Rz2z2}). 

The corresponding  statistics of the TWDP reference model given by  (\ref{Rzz1})-(\ref{Rz2z21}) are also included in the figures for comparison purpose, showing perfect match between simulated and theoretically obtained results in all cases except for the autocorrelation of the squared envelope in channel undergoing Rayleigh fading, as expected.  
Namely, in case when the strength of the diffuse component is large within the overall signal strength (i.e. when $K$ is close to 0), the term $4\sigma^4[f_s(2\pi f_D \tau,N)+f_c(2\pi f_D \tau,N)]/(2\Omega^2)$ in (\ref{Rz2z22}) becomes significant, producing notable difference between theoretical and simulated squared envelope autocorrelations.  
However, it is important to stress that the observed difference is the consequence of the model chosen to simulate diffuse component and can be reduced by using the one with more favorable characteristics. However, the scope of this paper was not to find the most accurate nor the most computationally efficient simulator. So underlying Rayleigh simulator is chosen from~\cite{Xia06} mostly due to its simplicity and the fact that it has published results related to correlation functions obtained for  Rician channels, which provide a reference point for verification of the results obtained using our simulator in case when TWDP fading collapses to Rician or Rayleigh.
Accordingly, obtained results given in Figs. (\ref{Fig_Rxx}) - (\ref{Fig_Rz2z2}) for Rayleigh and Rician fading are compared with those presented in~\cite[Figs. 3 - 5]{Xia06}, showing a perfect match between all curves obtained for the same set of parameters. 

\subsection{Evaluation of PDF of the Envelope}
\begin{figure}[t]
    \vspace{-0.4cm}
	\centering        \includegraphics[width=.425\textwidth]{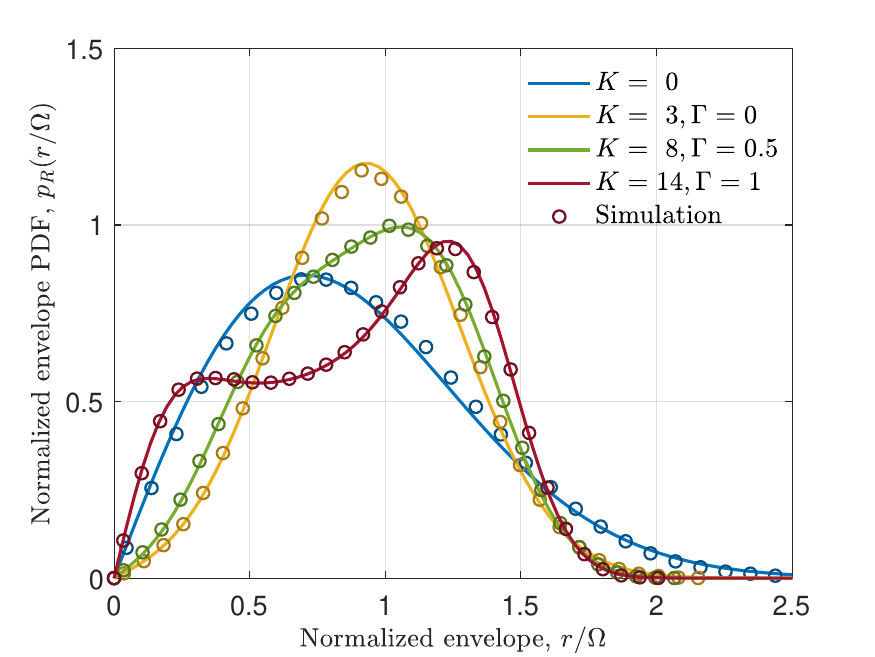}
	\caption{TWDP envelope PDF ($\alpha_1=\pi/4$, $\alpha_2=2\pi/3$)}
	\label{PDF}
\end{figure}

Fig.~\ref{PDF} shows the fading envelope PDF obtained using the proposed TWDP simulator with the specified set of simulation parameters, along with the results from the analytical expression for the TWDP envelope PDF given in\cite{Mar21}. Although better agreement between the simulated and theoretical results can be achieved by increasing $N>8$, the figure shows that the results obtained using proposed simulator agree very well with the theoretical ones, even for a small number of sinusoids ($N=8$). 

\begin{figure}[t]
    \vspace{-0.4cm}
	\centering        \includegraphics[width=.425\textwidth]{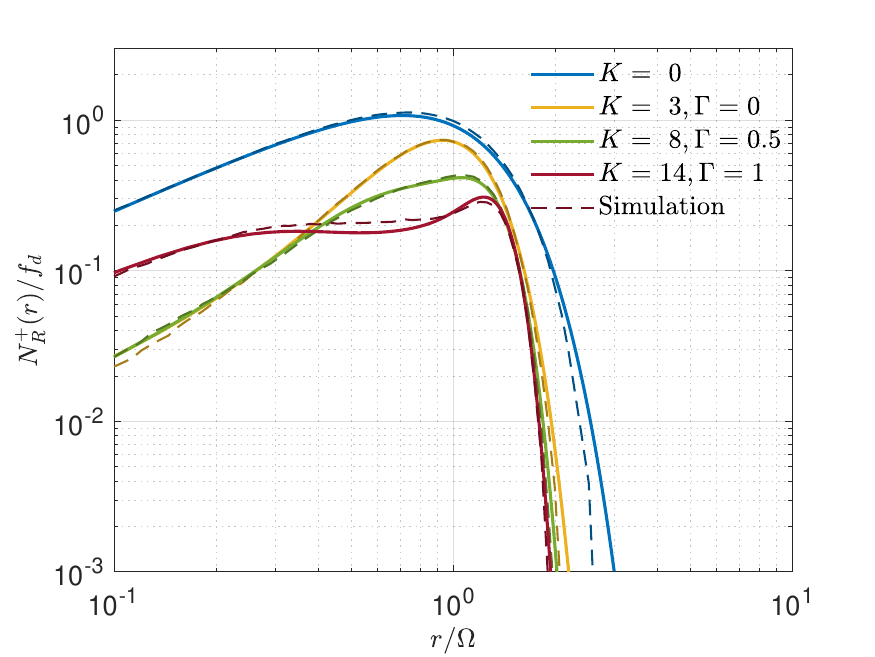}
	\caption{TWDP LCR ($\alpha_1=\pi/2$, $\alpha_2=-\pi/2$)}
	\label{LCR1}
\end{figure}

\begin{figure}[t]
    \vspace{-0.4cm}
	\centering        \includegraphics[width=.425\textwidth]{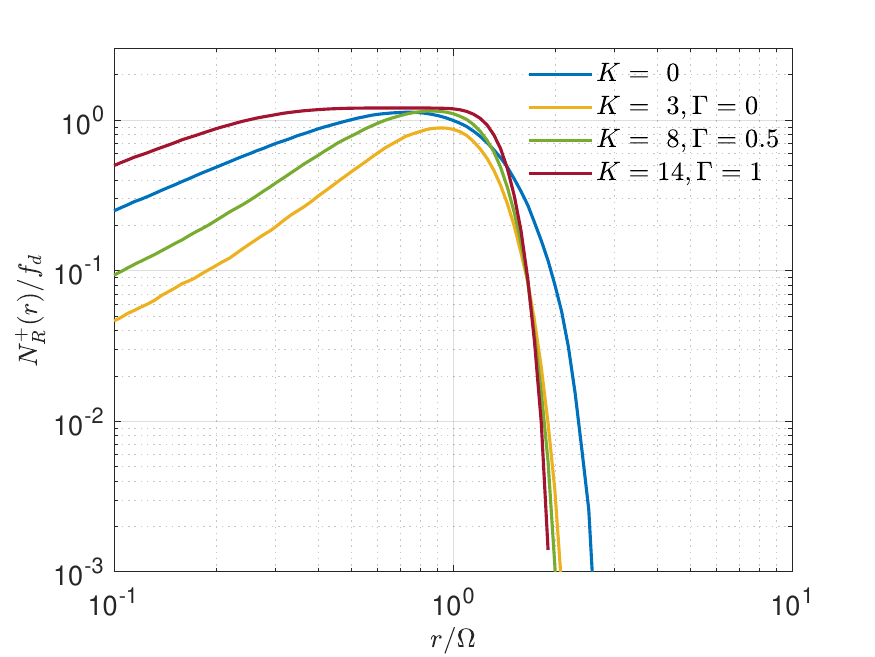}
	\caption{Simulated TWDP LCR, ($\alpha_1=\pi/4$, $\alpha_2=2\pi/3$)}
	\label{LCR2}
\end{figure}

\subsection{Evaluation of LCR}
The simulation results of the normalized level crossing rate (LCR) for the proposed TWDP simulator are shown in Figs.~\ref{LCR1} and~\ref{LCR2}. In Fig.~\ref{LCR1}, the results obtained using the proposed simulator are presented alongside those calculated using the analytical expression~\cite[(36)]{Rao14} derived for specific combination of AoAs in TWDP channels (i.e. when $\alpha_1=90^o$ and  $\alpha_2=-90^o$), showing excellent agreement in all considered cases. 
However, the expression~\cite[(36)]{Rao14} can only be used to obtain the LCR when both specular components arrive perpendicular to the direction of motion. For all other cases, no tools have been published to date for LCR evaluation. Therefore, the proposed simulator is the first tool that enables LCR evaluation for arbitrary TWDP parameters and AoAs.

For demonstration purpose, LCRs obtained using proposed simulator are plotted in Fig.~\ref{LCR2}, for different TWDP parameters and arbitrary AoAs, showing substantially different behaviors compared to the perpendicular cases presented in Fig.~\ref{LCR1}.
\section{Conclusion}
In this paper, the SoS-based TWDP channel simulator is proposed, enabling us to simulate Rayleigh, better-than-Rayleigh and worse-than-Rayleigh fading conditions. The simulator is constructed by using two zero-mean stochastic sinusoid with pre-selected Doppler frequencies and random initial phases, while the diffuse component is modeled using one of the existing Rayleigh channel simulators. For the described channel, analytical expressions for the autocorrelation and cross-correlation functions of quadrature components, as well as the autocorrelation of the complex envelope and its square, are first derived for the reference TWDP model. These expressions are then compared with those obtained using the proposed simulator for various combinations of channel and environmental parameters, demonstrating excellent agreement between them. Additionally, the proposed simulator is used to obtain diagrams of the envelope PDF and the LCR, which closely match those calculated using the existing analytical expressions. 
Since Rayleigh and Rician models are spatial cases of TWDP model itself, the results obtained using the proposed simulator (for specific sets of TWDP parameters) are finally compared to those from the literature obtained using the existing Rician/Rayleigh simulators, also showing the perfect match between them.

Accordingly, for the first time in the literature, a simulator is provided for accurate evaluation 
 of first- and second-order statistics of signals propagating in channels with TWDP fading. As such, it can serve as a valuable tool for optimizing the design of  interleaver/deinterleaver and channel coding/decoding units in these channels, enabling ultimate enhancement the overall performance of wireless communication systems operating in such environments.

\bibliographystyle{IEEEtran}

\begin{thebibliography}{10}
\providecommand{\url}[1]{#1}
\csname url@samestyle\endcsname
\providecommand{\newblock}{\relax}
\providecommand{\bibinfo}[2]{#2}
\providecommand{\BIBentrySTDinterwordspacing}{\spaceskip=0pt\relax}
\providecommand{\BIBentryALTinterwordstretchfactor}{4}
\providecommand{\BIBentryALTinterwordspacing}{\spaceskip=\fontdimen2\font plus
\BIBentryALTinterwordstretchfactor\fontdimen3\font minus \fontdimen4\font\relax}
\providecommand{\BIBforeignlanguage}[2]{{%
\expandafter\ifx\csname l@#1\endcsname\relax
\typeout{** WARNING: IEEEtran.bst: No hyphenation pattern has been}%
\typeout{** loaded for the language `#1'. Using the pattern for}%
\typeout{** the default language instead.}%
\else
\language=\csname l@#1\endcsname
\fi
#2}}
\providecommand{\BIBdecl}{\relax}
\BIBdecl

\bibitem{Xia06}
C.~Xiao, Y.~R. Zheng, and N.~C. Beaulieu, ``Novel sum-of-sinusoids simulation models for {R}ayleigh and {R}ician fading channels,'' \emph{IEEE Transactions on Wireless Communications}, vol.~5, no.~12, pp. 3667--3679, 2006.

\bibitem{Jak94}
W.~C. Jakes and D.~C. Cox, \emph{Microwave Mobile Communications}.\hskip 1em plus 0.5em minus 0.4em\relax Wiley-IEEE Press, 1994.

\bibitem{Pop01}
M.~Pop and N.~Beaulieu, ``Limitations of sum-of-sinusoids fading channel simulators,'' \emph{IEEE Transactions on Communications}, vol.~49, no.~4, pp. 699--708, 2001.

\bibitem{Zhe02}
Y.~Zheng and C.~Xiao, ``Improved models for the generation of multiple uncorrelated {R}ayleigh fading waveforms,'' \emph{IEEE Communications Letters}, vol.~6, no.~6, pp. 256--258, 2002.

\bibitem{Ali09}
A.~Alimohammad, S.~Fard, B.~Cockburn, and C.~Schlegel, ``Compact {R}ayleigh and {R}ician fading simulator based on random walk processes,'' \emph{Communications, IET}, vol.~3, pp. 1333 -- 1342, 09 2009.

\bibitem{Wan12}
J.~Wang, X.~rong Ma, J.~Teng, and Y.~Cui, ``Efficient and accurate simulator for {R}ayleigh and {R}ician fading,'' \emph{Transactions of Tianjin University}, vol.~18, pp. 243 -- 247, 2012.

\bibitem{Cla68}
R.~H. Clarke, ``A statistical theory of mobile-radio reception,'' \emph{The Bell System Technical Journal}, vol.~47, no.~6, pp. 957--1000, 1968.

\bibitem{Ski14}
M.~A. Skima, H.~Ghariani, and M.~Lahiani, ``A multi-criteria comparative analysis of different {R}ayleigh fading channel simulators,'' \emph{AEU - International Journal of Electronics and Communications}, vol.~68, no.~6, pp. 550--560, 2014.

\bibitem{Dur02}
G.~Durgin, T.~Rappaport, and D.~de~Wolf, ``New analytical models and probability density functions for fading in wireless communications,'' \emph{IEEE Transactions on Communications}, vol.~50, no.~6, pp. 1005--1015, 2002.

\bibitem{Pat98}
M.~Patzold and F.~Laue, ``Statistical properties of {J}akes' fading channel simulator,'' in \emph{VTC '98. 48th IEEE Vehicular Technology Conference}, vol.~2, 1998, pp. 712--718 vol.2.

\bibitem{Pat06}
M.~Pätzold and B.~Hogstad, ``Classes of sum-of-sinusoids {R}ayleigh fading channel simulators and their stationary and ergodic properties-{P}art {I},'' \emph{WSEAS Transactions on Mathematics}, vol.~5, pp. 222--230, 02 2006.

\bibitem{Pat05}
C.~Patel, G.~Stuber, and T.~Pratt, ``Comparative analysis of statistical models for the simulation of {R}ayleigh faded cellular channels,'' \emph{IEEE Transactions on Communications}, vol.~53, no.~6, pp. 1017--1026, 2005.

\bibitem{Mar21}
A.~Maric, E.~Kaljic, and P.~Njemcevic, ``An alternative statistical characterization of {TWDP} fading model,'' \emph{Sensors}, vol.~21, no.~22, 2021.

\bibitem{Rao14}
M.~Rao, F.~J. Lopez-Martinez, and A.~Goldsmith, ``Statistics and system performance metrics for the two wave with diffuse power fading model,'' in \emph{2014 48th Annual Conference on Information Sciences and Systems (CISS)}, 2014, pp. 1--6.

\end{thebibliography}

	





\end{document}